\documentstyle[12pt]{article}
\begin{document}
\newcommand{\be}{\begin{equation}}
\newcommand{\ee}{\end{equation}}

\begin{center}

{\bf Jet multiplicities as the QGP thermometer}

\vspace{2mm}

I.M. Dremin\footnote{e-mail: dremin@lpi.ru}, 
O.S. Shadrin\footnote{e-mail: oleg\_virt@bk.ru}

{\it Lebedev Physical Institute RAS, 119991 Moscow, Russia}

\end{center}

\vspace{1mm}

\begin{abstract}
It is proposed to use the energy behavior of mean multiplicities of jets
propagating in a nuclear medium as the thermometer of this medium during the
collision phases. The qualitative effects are demonstrated in the framework 
of the fixed coupling QCD with account of jet quenching.
\end{abstract}


\section{Introduction}

The properties of quark and gluon jets are firmly established and carefully
studied in $e^+e^-,\; ep$ and hadron collision experiments. They are well 
described by quantum chromodynamics (QCD) equations (see, e.g., reviews 
\cite{dr, koch, dgar}). The jets in $e^+e^-$ annihilation evolve in the 
QCD vacuum as a cascade process by stretching and breaking strings between 
color charges. The high energy behavior of mean jet multiplicities is uniquely 
determined by the QCD coupling strength $\alpha _S$. In the leading perturbative 
approximation it looks like 
\be
\langle n\rangle \propto \exp (\int ^{0.5\ln s} {\sqrt {2N_c\alpha _S(y')/\pi}dy'}).  \label{mult}
\ee

In case of high energy nucleus-nucleus collisions the similar hard jets 
propagate in a medium with properties different from those of vacuum. Between 
confined initial and final states, the system can pass the deconfined state of 
color charges (QGP) characterized by various temperature regimes. During its 
motion, the jet initiated by a high energy parton meets different (local) 
temperature conditions from cold to hot phases. Thus the energy (distance) 
scale behavior of the effective coupling strength differs from the traditional 
$1/\ln s$-law. In particular, it does not increase at large distances but 
flattens off and even decreases there at high temperatures. This has been shown 
by lattice calculations for heavy quarks in \cite{kzan, kkpz, nsom} (see also 
\cite{sore} and Fig. 3 below). Therefore, with the temperature-dependent 
coupling, mean multiplicities of jets can be used for measuring the QGP 
temperature as follows from (\ref{mult}).

For qualitative estimates, we will use two simplifying proposals. According 
to the lattice results, the energy (distance) scale dependence of the coupling 
strength becomes quite mild. It depends however on the temperatures. 
Averaging over different temperature regimes would flatten it even 
stronger. Therefore, in first approximation, it can be effectively 
replaced by a temperature-dependent constant at the distances 
most important for jet evolution. It simplifies the treatment of 
QCD integrodifferential equations for the generating functions of jet 
multiplicities because in this case they possess the scaling property 
and can be reduced to the system of algebraic equations which are exactly 
solved \cite{dhwa}.

Another well known property of "in-medium" jets which should be taken into 
account is jet quenching. Several different theoretical approaches to the
non-Abelian radiative energy loss of partons in dense QCD matter have been
proposed \cite{bdmp, glv, wied, wang}. For our qualitative estimates we use
the QCD interpretation of the medium-induced modification of single inclusive 
hadron spectra for such jets which was recently considered in \cite{bwie}. The 
kernels of the QCD equations were distorted so that the role of soft emissions 
was enhanced by introducing some new parameter in the infrared terms. This 
can be interpreted as the phenomenological description of the experimental 
fact of softening of jets spectra in nucleus-nucleus collisions. It implies 
that only soft partons with long wavelengths feel the neighboring deconfined 
partons of the medium surrounding the jets and multiplicate by rescattering. 
The leading perturbative terms of the QCD kernels play now even more important 
role. From the theoretical point of view it could be considered as a result of 
some yet unknown effective lagrangian which would be responsible for processes 
in the complicated nuclear medium and, probably, unite the above approaches. 
It accounts for the fact that jets lose their energy spending a part of it 
in the strong field of the surrounding quark-gluon matter.

Thus it is assumed that the hadronic medium changes both the coupling 
strength (the vertices) and the functional behavior of the kernels of QCD
equations. We show how these two modifications of QCD equations influence mean
multiplicities of hard jets piercing through the nucleus. This can be used
for measuring the medium temperature.

\section{QCD equations and their solution}

Any moment of the parton multiplicity distribution $P_n$ of gluon and quark 
jets in QCD can be obtained from the equations for the generating functions
\begin{eqnarray}
&G_{G}^{\prime }&= \int_{0}^{1}dxK_{G}^{G}(x)\gamma _{0}^{2}[G_{G}(y+\ln x)G_{G}
(y+\ln (1-x)) - G_{G}(y)] \nonumber \\
&+&n_{f}\int _{0}^{1}dxK_{G}^{F}(x)\gamma _{0}^{2}
[G_{F}(y+\ln x)G_{F}(y+\ln (1-x)) - G_{G}(y)] ,   \label{50}
\end{eqnarray}
\begin{equation}
G_{F}^{\prime } = \int _{0}^{1}dxK_{F}^{G}(x)\gamma _{0}^{2}[G_{G}(y+\ln x)
G_{F}(y+\ln (1-x)) - G_{F}(y)] ,                                   \label{51}
\end{equation}
where  $G_i$ are the generating functions of the multiplicity distributions
$P_n^{(i)}$ of gluon ($i=G$) and quark ($i=F$) jets, defined by
\begin{equation}
G_i(y,z)=\sum _{n=0}^{\infty }(z+1)^nP_n^{(i)}=\sum _{q=0}^{\infty }
\frac {z^q}{q!}\langle n_i\rangle ^qF_q^{(i)}.   \label{gen}
\end{equation}
In these expressions and below,  
$\langle n_i\rangle =\sum _{n=0}^{\infty }nP_n^{(i)}$ is the average
multiplicity, $z$ is an auxiliary variable, $y=\ln (p\Theta/Q_0)$ is the
evolution variable, $p, \Theta $ are the momentum and the opening angle of a
jet, $Q_0$=const ,
 $G^{\prime }(y)=dG/dy $, $ n_f$ is the number of active flavors,
 $N_c$ is the number of colors, 
and $C_{F}=  (N_{c}^{2}-1)/2N_{c}=4/3$ in QCD. 
 
 According to the above discussion, the coupling strength 
\begin{equation}
\alpha _S=\frac{\gamma _{0}^{2}\pi }{2N_{c}}      \label{alp}
\end{equation}
is set to be energy independent. Its temperature dependence will be however
crucial for further consideration. 

The modified kernels of the equations have been chosen similar to those
used in \cite{bwie}. They differ from the common QCD kernels \cite{dgar}
by the nuclear QCD factor $N_s$ which replaces 1 in soft infrared parts of the 
kernels. It changes the relative role of soft and hard splittings of partons.
\begin{equation}
K_{G}^{G}(x) = \frac {N_s}{x} - (1-x)[2-x(1-x)] ,    \label{53}
\end{equation}
\begin{equation}
K_{G}^{F}(x) = \frac {1}{4N_c}[x^{2}+(1-x)^{2}] ,  \label{54}
\end{equation}
\begin{equation}
K_{F}^{G}(x) = \frac {C_F}{N_c}\left[ \frac {N_s}{x}-1+\frac {x}{2}\right] .
\label{55}
\end{equation}
For $N_s=1$, they reduce to QCD kernels used for jets in $e^+e^-$ annihilation.
For $N_s>1$, the role of soft gluon emissions is enhanced (therefore, the label
$s$) that accounts for
their rescattering and multiplication in the nuclear medium of deconfined
partons. We do not show explicitely the $N_s$-dependence in the kernels' 
arguments.

The normalized factorial moment of any rank $q$ can be obtained by
differentiation
\begin{equation}
F_q^{(i)}=\frac {1}{\langle n_i\rangle ^q}\frac {d^qG_i}{dz^q}\vert _{z=0}, \label{fq}
\end{equation}
or, equivalently, by using the series (\ref {gen}) and collecting the terms
with equal powers of $z$ on both sides of the equations (\ref{50}), (\ref{51}).
In particular, the equations for mean multiplicities look like
\begin{eqnarray}
\langle n_G(y)\rangle ^{'} =\int dx\gamma _{0}^{2}[K_{G}^{G}(x)
(\langle n_G(y+\ln x)\rangle +\langle n_G(y+\ln (1-x)\rangle -\langle n_G(y)
\rangle ) \nonumber  \\
+n_{f}K_{G}^{F}(x)(\langle n_F(y+\ln x)\rangle +\langle n_F(y+
\ln (1-x)\rangle -\langle n_G(y)\rangle )],  \label{ng}
\end{eqnarray}
\begin{equation}
\langle n_F(y)\rangle ^{'} =\int dx\gamma _{0}^{2}K_{F}^{G}(x)
(\langle n_G(y+\ln x)\rangle +\langle n_F(y+\ln (1-x)\rangle -\langle n_F(y)
\rangle ).   \label{nq}
\end{equation}

Their solutions can be looked for \cite{dhwa} as
\be
\langle n_G\rangle \propto \exp (\gamma y), \;\; 
\langle n_F\rangle \propto r^{-1}\exp (\gamma y),  \label{ny}
\ee
i.e. the increase in multiplicities with energy follows a power law
$\langle n\rangle \propto s^{\gamma /2}$.
Both $\gamma $ and $r$ are energy independent. They depend however
on the temperature because $\alpha _S$ (\ref{alp}) depends on it and on
$N_s$-factor in the kernels.

The integrodifferential equations (\ref{50}), (\ref{51}) become the
algebraic equations
\be
\gamma=\gamma_0^2[M_1^G+n_f(r^{-1}M_1^F-M_0^F)],  \label{a1}
\ee
\be
\gamma=\gamma_0^2(L_2-L_0+rL_1).   \label{a2}
\ee
Here
\be
M_1^G=\int _0^1dxK_G^G[x^{\gamma }+(1-x)^{\gamma }-1], \nonumber
\ee
\be
M_1^F=\int _0^1dxK_G^F[x^{\gamma }+(1-x)^{\gamma }], \nonumber
\ee
\be
M_0^F=\int _0^1dxK_G^F,   \nonumber
\ee
\be
L_1=\int_0^1dxK_F^Gx^{\gamma }, \nonumber
\ee
\be
L_2=\int_0^1dxK_F^G(1-x)^{\gamma }, \nonumber
\ee
\be
L_0=\int_0^1dxK_F^G. \nonumber
\ee
Finally, these equations can be represented as the equation for $\gamma $
and the relation of $r$ to $\gamma $
\be
\left(\frac {\gamma }{\gamma_0^2}-a(\gamma )\right) 
\left(\frac {\gamma }{\gamma_0^2}-d(\gamma )\right)=b(\gamma )c(\gamma ), 
\label{gam}
\ee
\be
r(\gamma )=b(\gamma )\left(\frac {\gamma }{\gamma_0^2}-a(\gamma )\right)^{-1},
\label{rga}
\ee
where
\be
a=N_s(\psi(1)-\psi(\gamma+1)+B(\gamma ,1))-2B(\gamma +1,2)-2B(\gamma+2,1)+
B(\gamma+2,3)+B(\gamma+3,2)+\frac{11}{12}-\frac {n_f}{6N_c}, \nonumber
\ee
\be
b=\frac{n_f}{2N_c}[B(\gamma+3,1)+B(\gamma+1,3)],  \nonumber
\ee
\be
c=\frac {C_F}{N_c}[N_sB(\gamma,1)-B(\gamma+1,1)+0.5B(\gamma+2,1)], \nonumber
\ee
\be
d=\frac {C_F}{N_c}[N_s(\psi(1)-\psi(\gamma+1))-B(\gamma+1,1)+0.5B(\gamma+1,2)+0.75]. 
\nonumber
\ee
Here psi-functions and Euler beta-functions are used.

\section{Results and their discussion}

Eqs (\ref{gam}), (\ref{rga}) determine the experimentally measurable 
power in energy increase of jet multiplicities $\gamma $ and their ratio 
$r$ in gluon to quark jets as functions of the coupling strength $\alpha _S$ 
and the nuclear QCD factor $N_s$. 

Let us fix $\gamma $. Then both $\alpha _S$ and $r$ can be found 
from (\ref{gam}), (\ref{rga}) as functions of a single variable $N_s$. These
dependences are shown in Figs 1 and 2 for $\gamma = 0.4,\, 0.5,\, 0.6$.
The nearby values of $\gamma $ are chosen to show the sensitivity to 
$\alpha _S$ and $N_s$. They are grouped near $\gamma = 0.5$ simply because 
it is the only theoretically known power regime (obtained in relativistic
hydrodynamics). Other powers are admisssible in theory but, finally, the value
of $\gamma $ should be chosen according to experimental data. Our aim here is 
to demonstrate within this simplified model that, when measured, the definite 
values of $\gamma $ will reveal the temperature regimes of the matter during 
nuclear collisions, and, therefore, serve as a thermometer for the states of
this matter during the collison. This conclusion is however more general as 
follows from (\ref{mult}).

The effective range of $N_s$ was estimated in \cite{bwie} as changing from 1.6
to 1.8 for fits of the rapidity distributions. According to Fig. 1, it implies 
that the effective values of the coupling strength $\alpha _S$ are between 0.14 
and 0.122 for $\gamma = 0.5$. Then the average temperature of the regions inside 
nuclei which determine the jet evolution can be estimated by comparison of
these values with the lattice values of the coupling strength $\alpha _S$ 
calculated at different temperatures in \cite{kzan, kkpz, nsom} and presented 
also in \cite{sore}. To demonstrate this, the band
$0.122<\alpha _S<0.14$ is imposed in Fig. 3 on curves for distance dependence of
$\alpha _S$ at temperatures 3 $T_c$ and 6 $T_c$ (see Fig. 1 in \cite{sore},
and also Fig. 5 in \cite{kzan}, Fig. 2 in \cite{kkpz}). Herefrom, we conclude
that the effective temperature is surely above 3 $T_c$ and close to 6 $T_c$.
This would be the argument in favor of 
the statement that the jets penetrate through the deconfined state. The average 
values of $\alpha _S$ become smaller at higher temperatures because the strings 
are broken and the coupling strength decreases at large distances. 

This result seems quite reasonable for the chosen interval of $N_s$. The slower 
increase of multiplicities with energy would require (Fig. 1) too weak coupling 
strength ($0.07<\alpha _S<0.08$ for $\gamma =0.4$) and very high effective 
temperatures. The stronger increase can be hardly supported by experiment, and 
it would ask for stronger coupling ($0.22<\alpha _S<0.25$ for $\gamma =0.6$) 
and temperatures closer to $T_c$. Thus the energy dependence of mean 
multiplicities can be used for measuring the QGP temperature.

The values of the ratio of multiplicities in gluon to quark jets $r$ increase 
at larger $N_s$ as expected and seen from Fig. 2. They are larger at smaller 
$\gamma $. For $N_s\to \infty $ they tend to their asymptotical value 2.25 but 
are still quite far from it. In the region $1.6<N_s<1.8$ the ratio $r$ stays 
practically constant at fixed $\gamma $ and equal approximately 1.68, $\,$ 1.82,
$\,$ 1.95 at $\gamma $=0.6, $\,$ 0.5, $\,$ 0.4, correspondingly. We conclude 
that gluon jets are still more active in producing secondary partons in the
nuclear medium.

We have checked that we reproduce the limiting values of $\alpha _S$ and $r$ 
both for the "vacuum" QCD $N_s=1$ obtained in
\cite{dhwa} and for $N_s\to \infty$ which corresponds to the leading 
double-logarithmic approximation.

\section{Conclusions}

The nuclear medium impact on jet properties has been considered in QCD with 
fixed (but temperature-dependent) coupling. The enhancement of soft gluon 
emission (jet quenching) has been presented by the factor $N_s$ in the 
kernels of equations.

The general qualitative tendencies invoked by this factor $N_s$ for a 
constant power growth of the average multiplicities with energy increase
can be summarized in the following way:

1. The effective coupling strength $\alpha _S$ decreases with increase of $N_s$.

2. It is larger for larger power of multiplicity increase $\gamma $.

3. The ratio of gluon to quark jet multiplicities $r$ increases with increase 
of $N_s$.

4. It is larger for smaller power of multiplicity increase $\gamma $.

It follows from the first two statements that the energy increase of average 
multiplicities at definite values of $N_s$ can serve as the thermometer of 
the nuclear medium. The values of $N_s$ found in \cite{bwie} are larger than 1. 
It implies that $\alpha _S$ is lower than its zero-temperature value, i.e., 
in accordance with lattice results, the nuclear medium is characterized by 
the non-zero temperature. 

The gluon jets are still more active than quark jets in
producing secondary particles but this characteristics is not very sensitive
to the parameters of the medium.

The assumption of the fixed coupling can be relaxed inserting directly the 
lattice values for $\alpha _S$ at different temperatures in QCD equations. 
This complicates their computer solution. The work is in progress.

The behavior of the higher rank moments of multiplicity distributions and
nuclear modification of their oscillations (compared to jets in $e^+e^-$ 
annihilation studied in \cite{dhwa, 41}) can provide further insight in the
problem and will be described elsewhere.\\

{\bf Acknowledgements}

This work has been supported in part by the RFBR grants 03-02-16134, 
04-02-16445, NSH-1936.2003.2.\\

{\bf Figure captions.}

\begin{tabbing}
Fig. 1. \= The dependence of the coupling strength $\alpha _S$ on the nuclear \\ 
        \> QCD factor $N_s$ for $\gamma $=0.4 (bottom),$\,$ 0.5,$\,$ 0.6 (top).\\
Fig. 2. \> The dependence of the ratio of mean multiplicities in gluon \\
        \> and quark jets $r$ on $N_s$ for $\gamma $=0.4 (top),$\,$ 0.5,$\,$ 0.6 (bottom).\\
Fig. 3. \> The band of values $0.122<\alpha _S<0.14$ imposed on curves for \\
        \> distance dependence of $\alpha _S$ at temperatures 3 $T_c$ and 6 $T_c$  \\
        \> (the lattice results are taken from \cite{kzan, kkpz, nsom, sore}).\\
        \> The values for $T=0$ are shown by the curved line.\\
\end{tabbing}

\end{document}